\documentclass[12pt,openright]{article}
\usepackage{graphicx}
\usepackage{amssymb}		
\usepackage{lmodern}
\usepackage[]{algorithmicx}
\usepackage{algpseudocode}
\usepackage{sectsty}
\usepackage{setspace}
\usepackage{lineno}
\usepackage{amsmath}
\usepackage{graphicx}

\sectionfont{\fontsize{13}{15}\selectfont}
\subsectionfont{\fontsize{11}{15}\selectfont}
\usepackage{geometry}
\newtheorem{theorem}{Theorem}

\begin{document}
\title{\textbf{\large Layered graphs: a class that admits polynomial time solutions for some hard problems}
\date{\vspace{-5ex}}
\author{{\footnotesize $^{1,2}$Bhadrachalam Chitturi}}}
\maketitle
\begin{center}
{\footnotesize $^1$Department of Computer Science, Amrita Vishwa Vidyapeetham, Amrita University, Amritapuri Campus, Kollam, Kerala 690525, India.}\\
{\footnotesize $^2$Department of Computer Science, University of Texas at Dallas, Richardson, Texas 75083, USA.}
\end{center}

{\footnotesize\textbf{Abstract:}
The independent set on a graph $G=(V,E)$ is a subset of $V$ such that no two vertices in the subset have an edge between them. The MIS problem on $G$ seeks to identify an independent set with maximum cardinality, i.e. maximum independent set or MIS. $V* \subseteq V$ is a vertex cover $G=(V,E)$ if every edge in the graph is incident upon at least one vertex in $V*$. 
$V* \subseteq V$ is dominating set of $G=(V,E)$ if forall $v \in V$ either $v \in V*$ or $\exists u \in V*$ and $(u,v) \in E$. A connected dominating set, CDS, is a dominating set that forms a single component in $G$.
The MVC problem on $G$ seeks to identify a vertex cover with minimum cardinality, i.e. minimum vertex cover or MVC. Likewise, CVC seeks a connected vertex cover (CVC) with minimum cardinality.
The problems MDS and CDS seek to identify a dominating set and a connected dominating set respectively of minimum cardinalities.
MVC, CVC,  MDS, and CDS on a general graph are known to be NP-complete. 
On certain classes of graphs they can be computed in polynomial time. 
Such algorithms are known for bipartite graphs, chordal graphs, cycle graphs, comparability graphs, claw-free graphs, interval graphs and circular arc graphs for some of these problems.
In this article we introduce a new class of graphs called a layered graph and show that if the number of vertices in a layer is $O(\log \mid V \mid)$ then MIS, MVC, CVC, MDS and CDC can be computed in polynomial time. The restrictions that are employed on graph classes that admit polynomial time solutions for hard problems, e.g. lack of cycles, bipartiteness, planarity etc. are not applicable for this class.
}


\section{Introduction}
The maximum independent set problem on a graph $G=(V,E)$ seeks to identify a subset of $V$ with maximum cardinality such that no two vertices in the subset have an edge between them. If $V* \subseteq V$ is a  maximum independent set or MIS for short of $G$ then $\forall u,v \in V*,~ (u,v) \notin E$. In this article $G$ is undirected, so, an edge $(u,v)$ is understood to be an undirected edge. 

Karp proposed a method for proving problems to be NP-complete ~\cite{Karp1972}.  The maximum independent set problem on a general graph is known to be NP-complete \cite{Garey1979}. Certain classes of graphs admit a polynomial time solution for this problem. Such algorithms are known for trees and bipartite graphs \cite{Harary1969}, chordal graphs \cite{Gavril1972}, cycle graphs \cite{Gavril1973}, comparability graphs \cite{Golumbic1977}, claw-free graphs \cite{Minty1977}, interval graphs and circular arc graphs \cite{Gupta1982}. The maximum weight independent set problem is defined on a graph where the vertices are mapped to corresponding weights. The maximum weight independent set problem seeks to identify an independent set where the sum of the weights of the vertices is maximized.
On trees, the maximum independent set problem can be solved in linear time \cite{Chen1988}. Thus, for several classes of graphs MIS can be efficiently computed. \\
%
Hsiao et al. design an O(n) time algorithm to solve the maximum weight independent set problem on an interval graph with $n$ vertices given its interval representation with sorted endpoints list \cite{Hsiao1992}. Several articles improved the complexity of the exponential algorithms that compute an MIS on a general graph \cite{Tarjan1977, Robson1986}.  
Lozin and Milanic showed that MIS is polynomially solvable in the class of $S _{1, 2, k}$-free planar graphs, generalizing several previously known results where $S _{1, 2, k}$ is the graph consisting of three induced paths of lengths 1, 2 and $k$, with a common initial vertex \cite{Lozin2010}.

The minimum vertex cover problem on $G$ seeks to identify a vertex cover with minimum cardinality, i.e. minimum vertex cover or MVC. If $V* \subseteq V$ is MVC of $G$ then $\forall e=(u,v) \in E, u \in E \lor v \in E  $. In this article $G$ is undirected, so, an edge $(u,v)$ is understood to be an undirected edge. The minimum dominating set, i.e MDS and the minimum connected dominating set i.e. CDS problems seek to identify a dominating set and a connected dominating set respectively of minimum cardinality. The MVC, MDS and CDS probelms on a general graphs are known to be NP-complete \cite{Garey1979}. 
 Garey and Johnson showed that MVC is one first NP-complete problem~\cite{Garey1979}. In connected vertex cover (CVC) problem, given a connected graph G, we are required to find vertex cover set $C$ with minimum cardinality such that the induced subgraph G[C] is connected.
 Garey and Johnson proved that CVC is NP-complete~\cite{Garey1977}. For trees and bipartite graphs the minimum vertex cover can be identified in polynomial time~\cite{Bondy1976,Konig1931}. 
Garey and Johnson proved that CVC problem is NP-hard in planar graphs with a maximum degree of 4~\cite{Garey1979}. Li et. al. proved that for 4-regular graph CVC problem is NP-hard~\cite{Li2017}. It is shown that for series-parallel graphs, which are a set of planar graphs, it shown that minimum vertex cover can be computed in linear time~\cite{Takamizawa1982}.	
 
The minimum dominating set, i.e MDS and the minimum connected dominating set i.e. CDS problems seek to identify a dominating set and a  connected dominating set respectively of minimum cardinality. Garey and Johnson showed that MDS on planar graphs with maximum vertex degree 3 and planar graphs that are regular with degree 4 are NP-complete~\cite{Garey1979}. CDS is NP-complete even for planar graphs that are regular of degree 4~\cite{Garey1979}. Bertossi showed that the problem of finding a MDS is NP-complete for split graphs and bipartite graphs~\cite{Alan1984}. Cockayne et. al. proved that MDS in trees can be computed in linear time~\cite{Cockayne1975}. 
Haiko and Brandstadt showed that MDS and CDS are NP-complete for chordal bipartite graphs~\cite{Müller1987}. Ruo-Wei et. al. proved that for a given circular arc graph with $n$ sorted arcs, CDS is linear in time and space~\cite{Hung2002}. Fomin et. al. propose an algorithm with time complexity faster than $2^n$ for solving connected dominating set problem~\cite{Fomin2008}.
 
The term layered graph has been used in the literature. The hop-constrained  minimum spanning tree problem related to the design of centralized telecommunication networks with QoS constraints is  NP-hard \cite{Gouveia2011}. A graph that they call a \emph{layered graph} is constructed from the given input graph and authors show that  hop-constrained  minimum spanning  tree problem is equivalent to a Steiner tree problem. In software architecture the system is divided into several layers, this has been viewed as a graph with several layers. In this article we define a new class of graphs that we call \emph{layered graphs} and design an algorithm to identify the corresponding minimum vertex cover.

 \section{Layered Graph}
  
Consider a set of undirected graphs $G_1, G_2, \ldots G_p$ on the corresponding vertex sets $V_1, V_2, \ldots V_p$ and the edge sets $E_1, E_2, \ldots E_p$ i.e.  $G_i=(V_i, E_i)$. 
Consider a graph $G$ that is formed from $\forall_i~G_i$ with special additional edges called \emph{inter-layer edges} denoted as $E_{ij}$ where $j=i+1$ and $E_{ij}$ denotes the edges between $V_i$ and $V_j$. We call such a graph a \emph{layered graph} denoted as $LG$ whose layer number $i$ is $G_i$. Note that for any given $i$, $E_{ij}$ where $j=i+1$ can be $\phi$ and $\forall_{l \notin \{i-1,i+1\}} E_{il} =\phi$. Every vertex within a given layer gets a label from $(1,2, 3, \ldots, k)$. Thus, 
$ V_i \in \{ V_{i1}, V_{i2}, \ldots V_{ik} \} $.
 Note that $V_{ix}$ is the vertex number $x$ in layer $i$. However, in layer $i$ the vertex number $x$ need not exist. Further, if $(V_{ix}, V_{i+1~y}) \in E_{i~i+1}$ then it follows that vertex $x$ is present in layer $i$ and vertex $y$ is present in layer $i+1$.

We define the following restrictions on a layered graph. Several of the primary restrictions can be combined. Please see Figure 1.
\begin{itemize}
\item The size of all graphs is restricted such that $\mid V_i \mid \leq k$ then a \emph{k-restricted layered graph} i.e. $LG_{kr}$ is obtained. $LG_{kr}^q$ denotes an $LG$ with $q$ layers. $LG_{kr}^{q,n}$ denotes an $LG_{kr}^q$  with $n$ vertices.
\item If $\forall_{t}$ for $ V_{it}$ the only permissible edges are $(V_{it}, V_{jt})$  where $j \in \{ i-1, i+1\}$ then a \emph{linear layered graph} i.e. $LLG$ is obtained. $LLG_{kr}$ denotes an $LLG$ that is \emph{k-restricted}. $LLG_{kr}^q$ denotes an $LLG_{kr}$ with $q$ layers. $LG_{kr}^{q,n}$ denotes an $LLG_{kr}^q$  with $n$ vertices.
\item If every $G_i$ is required to be a connected component then a \emph{single component layered graph} i.e. $SLG$ is obtained.
\item If $G$ is required to be a connected component then a \emph{connected layered graph} i.e. $CLG$ is obtained.
\end{itemize}

This article primarily studies $LG_{kr}$ where every vertex within a given layer gets a label from $(1,2, 3, \ldots k)$. The results being general are applicable for $LLG$, $SLG$ and any combination thereof. Clearly, CDS, CVC are well defined only on $SLG$; further, the entire $G$ must be a single component. 
The recursive process of generating a hypercube of dimension $n+1$ i.e. $H_{n+1}$ from two copies of $H_{n}$
consists of creating the \emph{inter-}$H_{n}$ \emph{edges}  $\forall_i~(v_{1i},v_{2i} )$ where $v_{1i}$ and $v_{2i}$ are the corresponding vertices from the first copy of $H_{n}$ and the second copy of $H_{n}$ respectively. Thus, the \emph{inter-layer edges} of $LLG$ are in fact akin to a subset of \emph{inter-}$H_{n}$ edges because an \emph{inter-}$H_{n}$ edge exists between every pair of corresponding edges. However, in an $LLG$ the successive layers need not have all allowed edges; moreover, $\mid V_i\mid$ and $\mid V_{i+1}\mid$ need not be identical.

The complete graph on $k$ vertices is known as a clique on $k$ vertices and it is denoted by $K_k$. 
Consider a graph $G$ formed from several copies of $K_k$ say $G_1, G_2, \ldots G_p$ where in addition to the edges that exist in each of $G_i$ an edge is introduced between every pair $u, v$: $u \in G_i$ and $v \in G_{i+1}$.  We denote this particular graph $G$ that has $p$ layers with $K_k^p$. 
The class of \emph{k-restricted layered graphs} are in fact subgraphs of $K_k^p$. 
Thus, we call $K_k^p$ as \emph{full} $LG_{kr}^{p,n}$. Likewise, a $LLG$ that is defined on $p$ cliques where 
for any $i,i+1$ for all values of $l$ an edge is introduced between vertex $l$ of layer $i$ and vertex $l$ of layer $i+1$ is called as a \emph{full} $LLG_{kr}^{p,n}$.
The number of layers in $LG_{kr}$ i.e. $q$ is bounded by  $ n/k \leq q \leq n$. Likewise, the number of connected components in this graph lie in $[1, 2, 3 \ldots n]$. Recall that for any given $i$, $E_{i~i+1}$ can be $\phi$; that is the adjacent layers need not have any inter-layer edges among them. 

A subgraph of $G$ \emph{induced} by vertices $u_1, u_2, \ldots u_i$ consists of all vertices $u_1, u_2, \ldots u_i$ and all the edges defined on them. 
We design an algorithms that computes the cardinality of a MVC, CDS and CDS of any subgraph of $K_k^p$ i.e. $LG_{kr}^{q,n}$ with polynomial time complexity when $k= O(\log n)$. Moreover, we compute the number of MISs, MVCs, CDSs and CDSs in $LG_{kr}^{q,n}$. The worst case time complexity analysis which shows that these problems can be solved in polynomial time for $k= O(\log n)$ holds for all these problems even though the time complexities are not identical when the parameter $k$ is considered. 

CDS problem is meaningful only for $CLG$. However, CVC problem does not require $CLG$ because in the layers with no edges a vertex need not be selected. If every layer has at least one edge then CVC also requires $CLG$.
%
%
%
%
%
%
\section{Algorithm}
 Consider a layered graph with $q$ layers i.e. $LG_{kr}^{q, n}$ with layers $(1, 2, 3, \ldots, q)$.
 We design a generic dynamic programming algorithm for all the problems. The CDS is applicable only when $G$ is connected; thus, for these versions we are restricted to $CLG$.
 The specific details pertaining to each problem are elucidated. The layers are processed sequentially from the first to the last. Let $S=\bigcup_{j=1}^q V*_j $ be a candidate solution for a problem where $V*_j $ denotes the set of nodes that are chosen from layer $j$. 
 The candidate sub-solutions or simply $css$ for layer $i$ and layers $1 \ldots i$ are denoted as $css_i$ and $ccss_{i}$ respectively where $cccs$ denotes a combined candidate sub-solution for the first $i$ layers. 
 Likewise, $css_{i,j}$ and $ccss_{i,j}$ denote particular instances where the solution chosen in layer $i$ equals $j$ i.e. $j$ is a $k$-bit variable that we call mask that denotes the chosen vertices. 
 Further, $ccss_{i,j}$ potentially denotes several candidate sub-solutions of vertices from the first $i$ layers where $j$ is the mask corresponding to the layer $i$. We in general do not even store all these vertex sets but only the cardinality of the best option(s), such cardinality is called an \emph{optimum value}. This is stored in the variable $mask_{i,j}$ and the number of solutions that yield the optimum value when the mask for layer $i$ is $j$ is stored in $count_{i,j}$.\\ 
 The current layer requires the information only from the previous layer. So, only the variables of the current layer and the previous layer are maintained. The index 0 refers to the variables of the previous layer and index one refers to the same of the current layer. When the information for the index one is completely computed then it is overwritten onto the corresponding variables of index zero; the variables with index one are zeroed out and the next layer is processed (which becomes the current layer). This is a standard technique. Further, the overwriting of variables can be avoided by alternately assigning layers 1 and 0 as the current layer (the other becomes the corresponding previous layer).
 By employing extra space we can easily keep track of all possible solutions. However, we employ only $O(k 2^k)$ space (for domination problems we employ $O(k 2^{2k})$ space ) in addition to the space required by the graph. If we employ  $O(n 2^k)$ space ( for domination problems, $O(n 2^{2k})$ space) where for each mask in each layer we store a best compatible mask from the previous layer then we can additionally generate a solution.  However, if we want to generate all solutions then for each mask of a given layer we need to store all compatible masks of the previous layer that yield the optimum value. This further increases the space requirement. 
 
 We say that $css_{i,j}$ and $ccss_{i-1,l}$ are \emph{compatible} if $css_{i,j}\bigcup ccss_{i-1,l} \in ccss_{i,j}$. That is the union of $css_{i,j}$ and $ccss_{i-1,l}$ yields a $ccss$ for the first $i$ layers. Note that  compatibility is determined by  $css_{i,j}$ and $css_{i-1,l} \in ccss_{i-1,l}$ and the vertices chosen by $ccss_{i-1,l}$ in the earlier layers is irrelevant. This is a key feature. 
 
 \subsection{Input}
 The Input consists of $LG_{kr}^{q, n}$ that is specified in terms of $M_1, \ldots, M_q$ and $I_1, \ldots, I_{q-1}$ where $M_i$ is the 0-1 adjacency matrix for layer $i$ i.e. $G_i$. $I_i$ is the 0-1 adjacency matrix for $E_{i,i+1}$. 
  The rows $1,2, \ldots k$  of $I_i$ correspond to the vertices $V_{i1},V_{i2}, \ldots V_{ik}$  and the columns $1,2, \ldots k$ of $I_i$ are the  vertices $V_{i+1~1},V_{i+1~2}, \ldots V_{i+1~k}$. 
  It must be noted that for a linear graph $I_i$ can just be a $k$ dimensional vector and the corresponding computation is less expensive where $I_i[a]=1 \iff$ an edge between $a \in V_i, a \in V_{i+1}$ exists. 
  The boolean valued function \emph{compatible} is called to determine whether candidate sub-solutions (of the current layer and the subgraph induced up to the previous layer) can be combined; here the layer number $i$ is implicit. For each mask $j$ of a given layer $i$ a function $valid(i,j)$ determines if $j$ is a feasible option for layer $i$.
   The helper function $cardinality(j)$ returns the number of bits that are set in the binary representation of some mask $j$.

 All algorithms consist of the following sequence of computational tasks. 
 \begin{itemize}

 \item {Repeat (i) and (ii) for all layers $1 \ldots q-1$.}
 \item {(i) Feasible: $\forall_j$ (if $valid(i,j)$) then go to step(ii).}
  \item {(ii) Extension: If $j$ and $l$ are compatible then store the cardinality of $css_{i,j}\bigcup ccss_{i-1,l}$ in  $mask_{i,j}$ and the count of $ccss_{i,j}$ in $count_{i,j}$. Corresponding to each $css_{i,j}$ if $2^k$ additional variables are present then update them (e.g. DS problems).  
  }
  \item
 {(iii) Summarize: Layer $q:$  execute (i) and (ii). Identify the optimum cardinality among $\forall_j mask_{q,j}$ and the corresponding count.}
 \end{itemize}

Each problem has specific characteristics. The compatibility criteria and other specifics for each of the problems is elucidated below.

\subsection{MIS}
 Consider the structure of a MIS on $LG_{kr}^{q, n}$ say $ V^*= \bigcup_{j=1}^q V^*_j$ where $V^*_j$ are the vertices in MIS from layer $j$.  Clearly, $V^*_j$ must be an IS.
  Let $G_1$ be the subgraph of $LG_{kr}^{q, n}$ induced by $ V^1= \bigcup_{j=1}^i V_j$ and let $G_2$ be the subgraph of $LG_{kr}^{q, n}$ induced by $ V^2= \bigcup_{j=i+1}^q V_j$.  
 Consider the IS of $G$. IF $M_1 = \bigcup_{j=1}^i V^*_j $ and $M_2 =\bigcup_{j=1+1}^q V^*_j $ then $M_1$ and $M_2$ are ISs.
 Let the set of edges crossing the cut $C=(M_1, M_2)$ be $E^C$.
  It follows that the cardinality of an IS of $G$ is $ \mid M_1 \mid + \mid M_2 \mid$ when there is no edge crossing $C$. Note that the only edges that can go across the cut are $E_{i~i+1}$. Thus, the cardinality of MIS of $LG_{kr}^{q, n} =  max (\forall_{E^C= \phi} \mid M_1 \mid + \mid M_2 \mid)$.
  
  \begin{itemize}
   \item {$feasible(j)$: the mask $j$ must denote an IS for $G_i$.}
  \item {$compatible(j,l)$: the union of two ISs must be an IS.}
  \item {Extension: if$(cardinality(j) + mask_{i-1,l} > mask_{i,j})$ $mask_{i,j}\gets cardinality(j) + mask_{i-1,l}$.}
  \item{ Summarize: Let $opt \gets max(\forall_j mask_{q,j})$;$count\gets 0$;  $\forall_j~if(mask_{q,j} = opt) count \gets count+count_{q,j}$; Return $(opt, count_{q,j}))$
  		}
  \end{itemize}
  
\subsection{MVC and CVC} 
 Consider the VC $V^*=\bigcup_{j=1}^{j=q} V^*_i$ of $LG_{kr}^{q}$ where $V^*_i$ denotes the set of vertices in $V^*$ from layer $i$.  Clearly, $V^*_i$ is a VC for layer $i$.  $V^*_i$ depends only on $V^*_{i-1}$ and $V^*_{i+1}$. 
 
Consider two adjacent layers $p$ and $p+1$.  $V^*_{p}\bigcup V^*_{p+1}$ must cover all edges inter-layer edges between layers $p$ and $p+1$. Specifically,  $V^*=\bigcup_{j=1}^{j=p+1} V^*_i$ must cover all edges in the corresponding induced subgraph including $E_{p~p+1}$. Similar constraints hold for CMVC. Additionally the induced subgraph of $ V^*$ must be a single connected component.

 \begin{itemize}
   \item {$feasible(j)$: the mask $j$ must denote a VC for $G_i$. For CVC $j$ must be connected. }
  \item {$compatible(j,l)$: the union of two VCs must be a VC for edges in $G_i, G_{i+1}$ and $E_{i,j}$ .}
  \item {Extension: if$(cardinality(j) + mask_{i-1,l} < mask_{i,j})$ $mask_{i,j}\gets cardinality(j) + mask_{i-1,l}$. For CVC masks $j$ and $l$ must have at least on edge in between.} 
  \item{Summarize: Let $opt \gets min(\forall_j mask_{q,j})$;$count\gets 0$;  $\forall_j~if(mask_{q,j} = opt) count \gets count+count_{q,j}$; Return $(opt, count)$
  		}
  \end{itemize}
  
\subsection{MDS and CDS}
 Let the MDS on $LG_{kr}^{q, n}$ say $ V^*= \bigcup_{j=1}^q V^*_j$ where $V^*_j$ are the vertices in this MDS from layer $j$. Clearly, $V^*_j$ need not be a DS of layer $j$ because the $V_j$ can be dominated from any of $V^*_{j-1}, V^*_{j}, V^*_{j+1}$. It follows that $V^*_{p}\bigcup V^*_{p+1}$ must dominate all vertices in layers $p$. Specifically, the inclusion of $V^*_q $ in $ V^*$ must ensure the domination of $V_{q-1}\bigcup V_q$. 
 
 Consider $mask=j$ in layer $i$. Say, $css_{i,j}\bigcup ccss_{i-1,l}$ dominates layer $i-1$. However, this particular union of vertices leaves certain vertices in layer $i$ undominated. The number of such choices are $2^k$; each choice is denoted by a k-bit variable that we call mask, here, a mask of exclusion. Further, when one processes layer $i+1$ this information is significant. We show that $O(2^k)$ triples stored for each mask of a given layer suffice to compute MDS (CDC) of $LFG_{kr}$.
 %
 We aim to show that for a chosen mask $j$ in layer $i$ it suffices to store $2^k$ triples of the form $(m_x, s_x, c_x)$. Here $m_x$ is the mask of the vertices that are \emph{not} dominated in layer $i$, $s_x$ is the cardinality of the vertices chosen so far  and $c_x$ is the number of choices corresponding to $m_x$ for a particular $j$ in layer $i$. 
 The particular mask in the previous layer that is the cause for a particular triple in the current layer need not be carried forward. So, for MDS and CDS $mask_{i,j}$ indicates an array of $2^k$ triples $\forall_x (x,s_x,c_x)$ where $x$ is the mask of undominated vertices of layer $i$; $s_x,c_x$ are the respective size and count. 
 
 Let mask $j$ is chosen in layer $i$ it potentially be combined with every mask ( $O(2^k)$ masks) of the previous layer. Thus, potentially ($O(2^k)$) triples need be stored. Further, the total number of triples of the form $(m_x, s_x, c_x)$ is $\Omega(n.2^k)$ because $m_x$ can potentially assume any of $0 \ldots 2^k-1$, $s_x$ is $O(n)$ and $c_x$ can in fact be exponential in $n/k$. Here we make the following critical observations.
 \begin{itemize}
\item {Let the chosen mask for layer $i$ is $j$. When all the compatible vertex sets of the previous layer are considered then let the resultant triples for the choice of $j$ in layer $i$ be set $S$.  }
\item {In $S$ for any two triples with the same mask we need only retain the triples with the least size. The other triples cannot lead to an optimum solution.}
\item {If two triples have the same mask and the minimum size then they can be combined into one triple where the respective counts are added.}
\item {Thus, only $2^k$ triples suffice for a chosen mask for layer $i$. Which implies $2^{2k}$ triples suffice  $\forall_j~css(i,j)$. We store the information of only two layers. Thus, the space complexity is $O(k 2^{2k})$}

 \item {Thus, for a chosen mask for layer $i$ potentially $2^{2k}$ triples of previous layer must be processed. That is, for all masks of layer $i$, $2^{3k}$ triples of previous layer must be processed.
 		} 
 \item  { Consider the mask $j$ in layer $i$ and mask $l$ in layer $i-1$. Recall that there are $2^k$ triples stored
 			corresponding to mask $l$ in layer $i-1$. All the vertices that are covered by the combination of $j$ 
 			and $l$ in layer $i-1$ say $A$ and not covered in layer $i$ say $B$ can be computed in $O(k^2)$. This needs to be computed only once. Subsequently, for each triples stored corresponding to $l$ in layer $i-1$ we need only check if the undominated vertices are a subset of $B$ in $O(k)$ time.  Thus,  $O(k 2^k)$ is the dominating term in the time complexity yielding $O(k 2^{2k})$ for all masks of the previous layer. So, for all masks of the current layer the time complexity is $O(k 2^{3k})$. Thus, the time complexity of the algorithm is $O(n/k k 2^{3k})= O(n 2^{3k})$.
 		}
\end{itemize}
 
 
Similar constraints hold for CMDS. Additionally the induced subgraph of $ V^*$ must be a single connected component. Thus, $\forall_p  \bigcap V^*_p$ is connected. The time and space complexities are identical.
  
   \begin{itemize}
     \item {$feasible(j)$: For CMDS $j$ must be connected. For MDS any $j$ is valid.}
    \item {$compatible(j,l)$: the union must dominate all vertices of $V_{i-1}$. For CDC masks $j$ and $l$ must have at least on edge in between.}
    \item{Extension: Performed as per critical observations listed above.  The choice of the final layer must ensure that the final layer is dominated.}    
    \item{Summarize: Let $opt \gets min(\forall_j \forall_d size_{q,j,d} )$;$count\gets 0$; 
     $\forall_j \forall_d if (size_{q,j,d} = opt) count \gets count+ccount_{q,j,d}$; Return $(opt, count)$
    		}
    \end{itemize}
\subsection{Algorithm Compatible}
The function compatible receives two masks denoting chosen vertices from layers $i$ and $i+1$. If the vertices in layer $i+1$  dominate the so far undominated vertices in layer $i$ then the function returns true. Otherwise, it returns false. 
\vspace{4mm}
\noindent\textbf{Algorithm Compatible}
{\small
 \begin{algorithmic}[1]
  \State Input: $LG_{kr}$, $j$, $l$, and $I$.		 ~~~~~~~//The function call: $compatible(j, l)$. $j$: Mask for layer $i$. 
  \State Output: 0 (incompatible) or 1 (compatible). //$l$: Mask for layer $i+1$. $I$ denotes matrix for $E_{i~i+1}$.
  \State~~~~~~~~~~~~~~~~~~~~~~~~~~~~~~~~~~~~~~// $bit_c(i)$ returns true if bit $c$ is set in $i$ else returns false.
\newline  
  \State Case MIS:~~~~~~~~~~~~~~~~~~~~~~~~// Input: two valid MISs of two adjacent layers
  \If {$independent(j, l)$}~~~// $independent(j, l)$: for any $a,b:$ $bit_a(j)$ and $bit_b(l)$:  
  		\State {return 1;}~~~~~~~~~~~~~~~~~~~~~//if $I[a][b]=1$ return 0; otherwise return 1; $O(k^2)$ algorithm.
  		\Else
  		\State {return 0;}~~~~~~~~~~~~~~~~~~~//$\exists$ a pair of vertices across the layers joined with an edge.
  \EndIf
\newline  
   \State Case MVC:~~~~~~~~~~~~~~~~~~~~~~// Input: two VCs of two adjacent layers
    \If {$cover(j, l)$}~~~~~~~~~~~~// $cover(j, l)$: $\forall_{a,b}$ where $I[a][b]=1$: $a \land j >0$ or $b \land l >0$
    		\State {return 1;}~~~~~~~~~~~~~~~~~~~~// then return 1; otherwise return 0; $O(k^2)$ algorithm.
    		\Else
    		\State {return 0;}	
    \EndIf
\newline  
    \State Case CMVC:~~~~~~~~~~~~~~~~~~~~~~// Input: two CVCs of two adjacent layers
     \If {$ccover(j, l)$}~~~~~~~~~~~~~// $ccover(j, l)$: $\forall_{a,b}$ where $I[a][b]=1$: $a \land j >0$ or $b \land l >0$
     		\State {return 1;}~~~~~~~~~~~~~~~~~~~~~~// and $\exists_{c,d}:I[c][d]=1 \land bit_c(j) \land bit_d(l)$  							
     		\Else~~~~~~~~~~~~~~~~~~~~~~~~~~~~~~~~// then return 1; otherwise return 0; $O(k^2)$ algorithm.
     		\State {return 0;}
     \EndIf 
\newline
     \State Case MDS:~~~~~~~~~~~~~~~~~~~~~~// Input: two masks of two adjacent layers, 
      \If {$dom(j, l)$}~~~~~~~~~~~~~// $dom(j, l)$: $ D \leftarrow css_{i,j} \bigcup css_{i+1,l} \bigcup Adj(css_{i,j})  \bigcup Adj(css_{i+1,l})$
      		\State {return 1;}~~~~~~~~~~~~~~~~~~~// $i<q-1$: if $V_i \subseteq D$ then return 1; otherwise return 0;
      		\Else~~~~~~~~~~~~~~~~~~~~~~~~~~~~~// $i=q-1$: if $V_i \bigcup V_{i+1} \subseteq D$ then return 1; otherwise return 0; 
      		\State {return 0;}~~~~~~~~~~~~~~~~~~//$V_i$ or $V_i \bigcup V_{i+1}$ is not dominated. $O(k^2)$ algorithm.	
      \EndIf~~~~~~~~~~~~~~~~~~~~~~~~~// $Adj(V)$ is the set of all vertices neighboring any vertex in $V$
\newline
     \State Case CDS:~~~~~~~~~~~~~~~~~~~// Input: two masks of two adjacent layers, 
     \State~~~~~~~~~~~~~~~~~~~~~~~~~~~~~~~~//	$\exists_{c,d}:I[c][d]=1 \land bit_c(j) \land bit_d(l)$
      \If {$dom(j, l)$}~~~~~~~~~~// $dom(j, l)$: $ D \leftarrow css_{i,j} \bigcup css_{i+1,l} \bigcup Adj(css_{i,j})  \bigcup Adj(css_{i+1,l})$
      		\State {return 1;}~~~~~~~~~~~~~~~~// $i<q-1$: if $V_i \subseteq D$ then return 1; otherwise return 0;
      		\Else~~~~~~~~~~~~~~~~~~~~~~~~~// $i=q-1$: if $V_i \bigcup V_{i+1} \subseteq D$ then return 1; otherwise return 0; 
      		\State {return 0;}~~~~~~~~~~~~~~//$V_i$ or $V_i \bigcup V_{i+1}$ is not dominated. $O(k^2)$ algorithm.	
      \EndIf~~~~~~~~~~~~~~~~~~~~~// $Adj(V)$ is the set of all vertices neighboring any vertex in $V$
  
\end{algorithmic}
 }
\subsection{Algorithm Generic Optimum}
The algorithms for all the problems on $LG_{kr}$ is similar. We give a generic dynamic programming based algorithm. Some specific instances are shown in the Appendix.
Initialization: $\forall i~ mask_{0i}=mask_{1i}=0$;  $\forall i~ count_{0i}= count_{1i}= 0$;  $mask_{ij}:$ The minimum value of an vertex cover up to layer $i$ where the chosen vertices of the layer $i$ are given by the binary value of $j$. $count_{ij}:$ the number of ways the $j^{th}$ mask in layer $i$ yields the corresponding minimum value. The time complexity for MIS, MVC and CVC is $O(nk 2^{2k})$ and the space complexity is $O(k 2^k)$. The time complexity for MDS and CDS is $O(n 2^{3k})$ and the space complexity is $O(k 2^{2k})$.


\vspace{6mm}
 \noindent\textbf{Algorithm Generic Optimum}
 {\small
 \begin{algorithmic}
  \State Input: $LG_{kr}^{q, n}$
  \State Output: The cardinality and the count for the resp. problem.
 \For{ ($i=$0,...,$2^k-1$) }
   \If { $is(i)$ for layer 1}
  	\State { $count_{0i}=1; mask_{0i}=1;$ } // For all valid masks set their count
   \EndIf
 \EndFor
    
 \For{($i=2,....q$)}	//For layers 2 through maximum
 	\For {$(j=0,....2^k-1)$}  //For all masks of current layer
 		\State { Compose larger sub-solutions by considering all compatible masks of the}
 		\State { previous layer and any accompanying information.}
 	\EndFor				//Masks of previous layer	 		
 \EndFor				//For all layers
 
 \State $best \gets 0; sum \gets 0;$
 \For{(i=0,....$2^k-1$)}
  \State Identify $opt$, the cardinality of the optimum solution.
 \EndFor
 
 \For{($i=0,....2^k-1$)}
  \State Compute $count$, the count of optimum solutions.
 \EndFor
 
 \State $return(opt, count)$	
\end{algorithmic}
 }
 \section{Correctness and complexity}
 The correctness is shown for MIS and MVC problems. The time complexities are identical and the analysis is given for MVC. The proofs of correctness for the remaining problems are similar. The time complexity for domination problems was presented earlier.
 \begin{theorem}
  \label{MIScorrect}
   Algorithm correctly computes the MIS on $LG_{kr}^{q, n}$.
   \end{theorem}
  $\textbf{Proof:}$ Consider the structure of MIS on $LG_{kr}^{q, n}$.  Let $G_1$ be the subgraph of $LG_{kr}^{q, n}$ induced by $ V^1= \bigcup_{j=1}^i V_j$ and let $G_2$ be the subgraph of $LG_{kr}^{q, n}$ induced by $ V^2= \bigcup_{j=i+1}^q V_j$.  
  Consider the IS of $G$.  Let $M_1$ and $M_2$ be the independent sets of $G_1$ and $G_2$ respectively.
  Let the set of edges crossing the cut $C=(M_1, M_2)$ be $E^C$.
   It follows that the cardinality of an IS of $G$ is $ \mid M_1 \mid + \mid M_2 \mid$ when there is no edge crossing $C$. Note that the only edges that can go across the cut are $E_{i~i+1}$.  Thus, the cardinality of MIS of $LG_{kr}^{q, n} =  max (\forall_{E^C= \phi} \mid M_1 \mid + \mid M_2 \mid)$. This, is exactly being computed. The theorem follows. Another proof that can be given for this theorem is based on mathematical induction on subgraphs induced by the first $i$ layers. Likewise, $count_{ij}$ gives the number of ways an independent set of maximum cardinality that can be formed when the vertices chosen in the layer $i$ are given by $j$. Thus, $count_{qj}$ corresponding to the maximum value of  $mask_{qj}$ yields the total number of MISs.$\square$\\
 \begin{theorem}
 \label{MVCcorrect}
  Algorithm correctly computes the MVC on $LG_{kr}^{q, n}$.
 \end{theorem}
 $\textbf{Proof:}$ Consider the structure of MVC on $LG_{kr}^{q, n}$. Let $G_1$ be the subgraph of $LG_{kr}^{q, n}$ induced by $ V^1= \bigcup_{j=i}^i V_j$ and let $G_2$ be the subgraph of $LG_{kr}^{q, n}$ induced by $ V^2= \bigcup_{j=i+1}^q V_j$.  
 Consider the VC of $G$.  Let $M_1$ and $M_2$ be the vertex covers of $G_1$ and $G_2$ respectively.
 Let the set of edges crossing the cut $C=(M_1, M_2)$ be $E^C$.
  It follows that the cardinality of an VC of $G$ is $ \mid M_1 \mid + \mid M_2 \mid$ when every edge crossing $C$ is covered by either $M_1$ or $M_2$. Note that the only edges that can go across the cut are $E_{i~i+1}$.  Thus, the cardinality of MVC of $LG_{kr}^{q, n} =  min (\forall_{E^C= \phi} \mid M_1 \mid + \mid M_2 \mid)$. This, is exactly being computed. The theorem follows. Another proof that can be given for this theorem is based on mathematical induction on subgraphs induced by the first $i$ layers. Likewise, $count_{ij}$ gives the number of ways an vertex cover of minimum cardinality that can be formed when the vertices chosen in the layer $i$ are given by $j$. Thus, $count_{qj}$ corresponding to the minimum value of  $mask_{qj}$ yields the total number of MVCs. $\square$.\\
\begin{theorem}
\label{MVCinP}
 Algorithm MVC on $LG_{kr}^{q, n}$ runs in polynomial time in $n$ when $k=O(\log n)$. The space required is $O(k^2)$.
 \end{theorem}
$ \textbf{Proof:}$ We presume that $I_{i}$, the 0-1 adjacency matrix for the subgraph induced by $V_i \bigcup V_{i+1}$ where the edges are restricted to $E_{i~i+1}$ is given. Likewise, we assume that the 0-1 adjacency matrix $M_i$ for each of $G_i$ are given. Recall that $LG_{kr}^{q, n}$ was formed from $G_1, G_2, \ldots G_q$. For a linear graph, $I_i$ is just a $k-$dimensional vector where if bit $j$ is set then there is an edge between $V_{ij}$ and $V_{i+1~j}$.

\begin{itemize}
   \item {The initialization step requires $O(2^k)$ time.}
   \item {Given a mask for layer $i$ it can be determined if it is a valid VC in $O(k^2)$ time with $M_i$. That is, for any two $M_i[a][b]$ that is set the mask should have either bit $a$ or bit $b$ set. }
   \item {Given two masks $mask1, mask2$ for layers $i, i+1$ respectively and $I_i$ it can be directly determined if their union is a VC of a subgraph induced by $\bigcup_i^{i+1} V_j $ of $LG_{kr}^{q, n}$ in $O(k^2)$ time.}
   
   \item {In order to determine the MVC up to layer $i$ whose mask is $j$; $j$ must be checked for compatibility with all masks of the previous layer. Thus, $O(k^2 2^k)$ time is required. For all masks of the current layer $O(k^2 2^{2k})$ time is required. For all layers, the time required is maximized when each layer has $k$ vertices yielding  $O(n/k k^2 2^{2k})$ = $O(nk 2^{2k})$ time.}
 \end{itemize}
The time complexity is clearly exponential in $k$; however, if $k=O(1)$ the time complexity is $O(n)$. The time complexity remains polynomial when $k=O(\log n)$. The space required is  $O(k 2^k)$ because for two layers we store $4. 2^k$ mask and count variables each of size $k$. $\square$ \\
%
%


\section{Conclusions}
A new large class of graphs called layered graphs have been defined. These graphs can have exponential number of cycles.
This class includes a subset of bipartite graphs and a subset of trees on $n$ vertices.  The restrictions that are employed on graph classes that admit polynomial time solutions for hard problems, e.g. lack of cycles, bipartiteness, planarity etc. are not applicable for this class.
The computation of hard problems on these graphs along with the count of the corresponding optimum solutions is shown to be in $P$ when layer size is $O(\log \mid V \mid)$. 

\begin{figure}[h!]
\begin{center}
\includegraphics[]{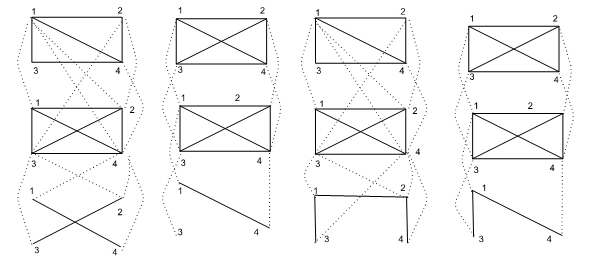}
\end{center}
\begin{center}
\caption{ From left to right: 1a) $LG_{4r}^{3,12}$. 1b) $LLG_{4r}^{3,10}$. 1c) $SLG_{4r}^{3,11}$. 1d) $SLLG_{4r}^{3,11}$.  
In single component graphs, each layer has exactly one connected component. The vertices are labeled $1,2,3,4$ within the given layer. The edges between the vertices of a given layer are shown with thick lines whereas an $e \in E_{i~i+1}$ is shown with a dotted line. The graph is labeled. In a linear graph the edges $ \in E_{i~i+1}$ connect the vertices with identical labels from adjacent layers.}
\end{center} 
\end{figure}



\section {Acknowledgments}
 Authors thank Amma for her guidance.  This article is based on technical report UTDCS-02-17.

\begin{figure}[h!]
\begin{center}
\includegraphics[]{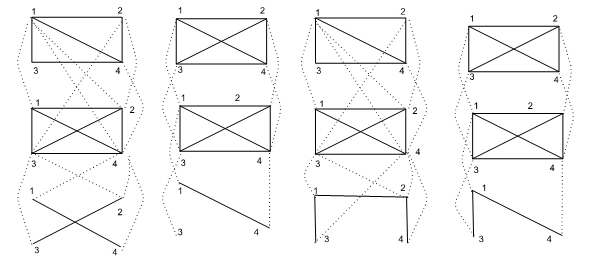}
\end{center}
\begin{center}
\caption{ From left to right: 1a) $LG_{4r}^{3,12}$. 1b) $LLG_{4r}^{3,10}$. 1c) $SLG_{4r}^{3,11}$. 1d) $SLLG_{4r}^{3,11}$.  
In single component graphs, each layer has exactly one connected component. The vertices are labeled $1,2,3,4$ within the given layer. The edges between the vertices of a given layer are shown with thick lines whereas an $e \in E_{i~i+1}$ is shown with a dotted line. The graph is labeled. In a linear graph the edges $ \in E_{i~i+1}$ connect the vertices with identical labels from adjacent layers.}
\end{center} 
\end{figure}
\section{Appendix}
The generic algorithm was presented earlier. Here, we state two specific algorithms for MIS and MVC respectively.\\\\
\noindent\textbf{Algorithm MIS}
 {\small
 \begin{algorithmic}
  \State Input: $LG_{kr}^{q, n}$
  \State Output: The cardinality of MIS and the count of the maximum independent sets.
  \State Initialization: $\forall i~ mask_{0i}=mask_{1i}=0$; 
  \State $\forall i~ count_{0i}= count_{1i}= 0$;  
  \State //$mask_{ij}:$ The maximum value of an independent set up to layer $i$ where the chosen 
  \State //vertices of the layer $i$ are given by the binary value of $j$.
  \State //$count_{ij}:$ the number of ways the $j^{th}$ mask in layer $i$ yields the corresponding maximum value.
  \State //$is(j)$ is a boolean function that returns true if the vertex assignment in the current layer
  \State //corresponding to the binary value of $j$ forms an IS. Otherwise it returns false.
  \State //$\land$ is the bitwise AND operator.
  \State //$cardinality(j)$ is the number of bits that are set in the binary representation of $j$.
  \State // For each $mask_{ij}$ one $k$-bit variable that remembers the mask of the layer $i-1$ that
  \State // yielded $mask_{ij}$ will help in constructing MISs. Union of such masks (1/layer) is an MIS.
  \State // This will require $O(n/k 2^k)$ additional space. Current version needs only $O(2^k)$ space. 
 \For{ ($i=$0,...,$2^k-1$) }
   \If { $is(i)$ for layer 1}
  	\State { $count_{0i}=1; mask_{0i}=1;$ } // No. of valid ISs of layer 1
   \EndIf
 \EndFor
    
 \For{($i=2,....q$)}	//For layers 2 through maximum
 
 	\For {$(j=0,....2^k-1)$}  //For all masks of current layer
 		\If {$is(j)$}		//$j$ is valid
 		\State $size \gets 0$
 		\For {$(l=0,...,2^k-1)$ }	//Masks of previous layer	
  			\If { $((mask_{0l}>0) \land (compatible(j,l)))$ }	//$mask_{0l}=0 \rightarrow$Invalid IS
   				\If  { $(cardinality(j) + mask_{0l} >size)$}	// Better IS for the current mask 
   					\State		{$size = cardinality(j)+mask_{0l};$}
   				\EndIf
    	   \EndIf
    		\State $mask_{1j} \gets size$
 		\EndFor				//Masks of previous layer	
 		
		\For {$(l=0,...,2^k-1)$} 	//Masks of previous layer	
  			\If {($size = cardinality(j)+mask_{0l};$)}					//Instance of max
   				\State {$count_{1j} \gets count_{1j} + count_{0l};$}	// Count corr. to max wrt mask=$j$
    		\EndIf
 		\EndFor				//Masks of previous layer

		\EndIf					// $j$ is valid
 	\EndFor				//For all masks of current layer
 	
 	\State $\forall x~ count_{0x} \gets count_{1x};  mask_{0x} \gets mask_{1x}; count_{1x} \gets mask_{1x} \gets 0$;
 
	\EndFor			//For layers 2 through maximum
 
 \State $best \gets 0; sum \gets 0;$
 \For{(i=0,....$2^k-1$)}
 		 \If {$mask_{1i} > best$}	//Get the max value of $\forall_i mask_{pi}$
 		 \State {$best = mask_{1i};$}
 		\EndIf
 \EndFor
 
 \For{($i=0,....2^k-1$)}
  		 \If {$mask_{1i}= best$}	//Corr. to the best value of $MIS(LG_{kr}^{q, n})$
  		 \State {$sum \gets sum + count_{1i};$}	//Get the count of MISs
  		\EndIf
  \EndFor
 
 \State $return(best, sum)$	//MIS cardinality and the count of such MISs
\end{algorithmic}
 }
%
%
%
%
 \vspace{6mm}
  \noindent\textbf{Algorithm MVC}
  {\small
  \begin{algorithmic}
   \State Input: $LG_{kr}^{q, n}$
   \State Output: The cardinality and the count for the resp. problem.
  \For{ ($i=$0,...,$2^k-1$) }
    \If { $is(i)$ for layer 1}
   	\State { $count_{0i}=1; mask_{0i}=1;$ } // No. of valid VCs of layer 1
    \EndIf
  \EndFor
     
  \For{($i=2,....q$)}	//For layers 2 through maximum
 
  	\For {$(j=0,....2^k-1)$}  //For all masks of current layer
  		\If {$is(j)$}		//$j$ is valid
  		\State $size \gets 0$
  		\For {$(l=0,...,2^k-1)$ }	//Masks of previous layer	
   			\If { $((mask_{0l}>0) \land (compatible(j,l)))$ }	//$mask_{0l}=0 \rightarrow$Invalid VC
    				\If  { $(cardinality(j) + mask_{0l} <size)$}	// Better VC for the current mask 
    					\State		{$size = cardinality(j)+mask_{0l};$}
    				\EndIf
     	   \EndIf
     		\State $mask_{1j} \gets size$
  		\EndFor				//Masks of previous layer	
  		
 		\For {$(l=0,...,2^k-1)$} 	//Masks of previous layer	
   			\If {($size = cardinality(j)+mask_{0l};$)}					//Instance of max
    				\State {$count_{1j} \gets count_{1j} + count_{0l};$}	// Count corr. to max wrt mask=$j$
     		\EndIf
  		\EndFor				//Masks of previous layer

 		\EndIf					// $j$ is valid
  	\EndFor				//For all masks of current layer
  	
  	\State $\forall x~ count_{0x} \gets count_{1x};  mask_{0x} \gets mask_{1x}; count_{1x} \gets mask_{1x} \gets 0$;
  
 	\EndFor			//For layers 2 through maximum
  
  \State $best \gets 0; sum \gets 0;$
  \For{(i=0,....$2^k-1$)}
  		 \If {$mask_{1i} < best$}	//Get the max value of $\forall_i mask_{pi}$
  		 \State {$best = mask_{1i};$}
  		\EndIf
  \EndFor
  
  \For{($i=0,....2^k-1$)}
   		 \If {$mask_{1i}= best$}	//Corr. to the best value of $MVC(LG_{kr}^{q, n})$
   		 \State {$sum \gets sum + count_{1i};$}	//Get the count of MVCs
   		\EndIf
   \EndFor
  
  \State $return(best, sum)$	//MVC cardinality and the count of such MVCs
 \end{algorithmic}
  }

\begin{thebibliography}{}

\bibitem{Harary1969}
Harary, Frank. Graph theory. 1969.

\bibitem{Gavril1972}
Gavril, Fanika. Algorithms for minimum coloring, maximum clique, minimum covering by cliques, and maximum vertex cover of a chordal graph. SIAM Journal on Computing 1.2 (1972): 180-187.
%
\bibitem{Gavril1973}
Gavril, Fanika. Algorithms for a maximum clique and a maximum independent set of a circle graph. Networks 3.3 (1973): 261-273.
%
\bibitem{Cockayne1975}
Cockayne, E., Sue Goodman, and Stephen Hedetniemi. A linear algorithm for the domination number of a tree. Information Processing Letters 4.2 (1975): 41-44.
%
\bibitem{Tarjan1977}
Tarjan, Robert Endre, and Anthony E. Trojanowski. Finding a maximum independent set. SIAM Journal on Computing 6.3 (1977): 537-546.
%
\bibitem{Golumbic1977}
Golumbic, Martin Charles. The complexity of comparability graph recognition and coloring. Computing 18.3 (1977): 199-208.
%
\bibitem{Minty1977}
Minty, George J. On maximal independent sets of vertices in claw-free graphs. Journal of Combinatorial Theory, Series B 28.3 (1980): 284-304.
%
\bibitem{Gupta1982}
Gupta, Udaiprakash I., Der-Tsai Lee, and JY‐T. Leung. Efficient algorithms for interval graphs and circular arc graphs. Networks 12.4 (1982): 459-467.
%
\bibitem{Robson1986}
Robson, John Michael. Algorithms for maximum independent sets. Journal of Algorithms 7.3 (1986): 425-440.
%
\bibitem{Chen1988}
Chen, Gen-Huey, M. T. Kuo, and J. P. Sheu. An optimal time algorithm for finding a maximum weight independent set in a tree. BIT Numerical Mathematics 28.2 (1988): 353-356.
%
\bibitem{Gavril2000}
Gavril, Fanica. Maximum weight independent sets and cliques in intersection graphs of filaments. Information Processing Letters 73.5-6 (2000): 181-188.
%
\bibitem{Hsiao1992}
Hsiao, Ju Yuan, Chuan Yi Tang, and Ruay Shiung Chang. An efficient algorithm for finding a maximum weight 2-independent set on interval graphs. Information Processing Letters 43.5 (1992): 229-235.
%
\bibitem{Lozin2010}
Vadim V. Lozin, and Martin Milanic. On the Maximum Independent Set Problem in Subclasses of Planar Graphs. Journal of Graph Algorithms and Applications 14.2 (2010): 269-286. 
%
\bibitem{Gouveia2011}
Gouveia, Luis, Luidi Simonetti, and Eduardo Uchoa. Modeling hop-constrained and diameter-constrained minimum spanning tree problems as Steiner tree problems over layered graphs. Mathematical Programming 128.1 (2011): 123-148.
%
\bibitem{Garey1979}
Garey, Michael R., and David S. Johnson. Computers and Intractability: A Guide to the Theory of NP-completeness. (1979).

\bibitem{Chitturi2017}
Bhadrachalam Chitturi. The maximum independent set problem on layered graphs. Technical Report: UTDCS-02-17, Department of Computer Science, University of Texas at Dallas, 2017.

\bibitem{Karp1972}
Karp, Richard M. Reducibility among combinatorial problems. Complexity of computer computations. 1972.

\bibitem{Garey1977}
Garey, M.R., D. S. Johnson, The rectilinear Steiner tree problem is NP-complete. SIAM J. Appl.Math. 826-834.

\bibitem{Li2017}
Li, Yuchao and Yang, Zishen and Wang, Wei. Complexity and algorithms for the connected vertex cover problem in 4-regular graphs. Applied Mathematics and Computation. 301(2017): 107-114.

\bibitem{Bondy1976}
Bondy, John Adrian, and Uppaluri Siva Ramachandra Murty. Graph theory with applications. London: Macmillan, 1976.

\bibitem{Konig1931}
Konig, Denes. Graphen und matrizen. Mat. Fiz. Lapok 38.1931(1931): 116-119.

\bibitem{Alan1984}
Bertossi, Alan A. Dominating sets for split and bipartite graphs. Information processing letters 19.1(1984): 37-40.

\bibitem{Takamizawa1982}
Takamizawa, Kazuhiko, Takao Nishizeki, and Nobuji Saito. Linear-time computability of combinatorial problems on series-parallel graphs. Journal of the ACM (JACM) 29.3(1982): 623-641.


\bibitem{Müller1987}
Muller, Haiko, and Andreas Brandstadt. The NP-completeness of Steiner tree and dominating set for chordal bipartite graphs. Theoretical Computer Science 53.2(1987): 257-265.

\bibitem{Hung2002}
Hung, Ruo-Wei, Maw-Shang Chang, and Chiayi Ming-Hsiung. A linear algorithm for the connected domination problem on circular-arc graphs. Proceedings of the 19th Workshop on Combinatorial Mathematics and Computation Theory. 2002.

\bibitem{Fomin2008}
Fomin, Fedor V., Fabrizio Grandoni, and Dieter Kratsch. Solving connected dominating set faster than $2^n$. Algorithmica 52.2(2008): 153-166.

\end{thebibliography}
\end{document}